\documentclass[runningheads]{llncs}
\usepackage{graphicx}
\usepackage{booktabs}       % professional-quality tables
\usepackage{amsfonts}       % blackboard math symbols
\usepackage{nicefrac}       % compact symbols for 1/2, etc.
\usepackage{microtype}      % microtypography
\usepackage{xcolor}         % colors
\usepackage{algorithm}
\usepackage{algpseudocode}
\usepackage{tikz} 
\usetikzlibrary{trees}
\usepackage{amsmath}
\usepackage{amssymb}
\makeatletter
\DeclareRobustCommand{\sqcdot}{\mathbin{\mathpalette\morphic@sqcdot\relax}}
\newcommand{\morphic@sqcdot}[2]{%
  \sbox\z@{$\m@th#1\centerdot$}%
  \ht\z@=.33333\ht\z@
  \vcenter{\box\z@}%
}
\makeatother

\newcommand{\scheme}[1]{\textsf{#1}}
\newcommand{\operation}[1]{\textsf{#1}}
\newcommand{\library}[1]{\textsc{#1}}

\newcommand{\feature}[2]{{#1}^{(#2)}}

\author{[~Anonymous submission to ThisML Conference 2022~]}

\authorrunning{J. Frery et al.}
\titlerunning{Privacy-Preserving Tree-Based Inference with FHE}

\begin{document}
\title{Privacy-Preserving Tree-Based Inference with TFHE}
\author{Jordan Frery \and Andrei Stoian \and  Roman Bredehoft \and Luis Montero \and 
Celia Kherfallah \and 
Benoit Chevallier-Mames \and Arthur Meyre}

\institute{Zama \footnote{
\email{hello@zama.ai} \url{http://zama.ai}} }

% \authorrunning{A. Stoian et al.}
\maketitle              % typeset the header of the contribution

%Each submission must be in English, in PDF format, and begin with a cover page including: (1) the title, (2) the names and affiliations of all authors, (3) contact author’s email and address, (4) a couple of paragraphs abstract of the paper, (5) indication whether the paper is a regular submission, or a short paper submission.

\begin{abstract}
Fully Homomorphic Encryption is a powerful tool for processing encrypted data and is particularly adapted to the type of programs that are common in machine learning (ML). On tabular data, tree-based ML models obtain state-of-the-art results, are more robust, and are easier to use and deploy than neural networks. We introduce an implementation of privacy-preserving decision tree evaluation based on the TFHE scheme, leveraging optimized representations for encrypted integer and TFHE's powerful programmable boostrapping mechanism. Our technique is applicable to decision trees, random forests, and gradient boosted trees. We demonstrate our approach on popular datasets and show that accuracy on encrypted data is very close the one obtained by the same models applied to clear data, while latency is competitive with the state of the art.
\end{abstract}

%\section*{CSCML 2023 Submission}

%This paper is submitted as a \textbf{long paper}.

%\newpage

\section{Introduction}

Over the past decade, machine learning has become a powerful tool in many applications such as  image classification, automatic translation, speech-to-text, voice synthesis, image and text generation. The rise of ML is owed to advancements in hardware and to higher data availability, which have allowed training of increasingly complex models. Additionally, the development of deep learning techniques, such as convolutional neural networks, has played a significant role in the success of ML in various domains.

Tree-based models are a popular choice for both classification and regression, and they are particularly adapted for tabular data. Indeed, studies have shown that tree-based models are still the best solution for this type of data, which contains categorical features that are usually expressed with highly sparse one-hot encodings ~\cite{grinsztajn2022tree}. Furthermore, tree-based models are scale-invariant and, thus, are easy to train as they obviate specific feature engineering or pre-processing.

Machine learning models are often applied to sensitive data in use-cases such as facial recognition, biometrics, finance, advertising and healthcare. Due to data confidentiality issues , processing health data to gain insights or to diagnose diseases using artificial intelligence lags behind, in terms of adoption. For example, most hospitals have restrictions on sharing patient data with third parties, which often prevents them from using  state-of-the-art data-driven algorithms. Privacy-preserving machine learning (PPML) inference on encrypted data would be a powerful tool to protect the privacy of data while still allowing for accurate predictions \cite{al2019privacy}. Several types of approaches exist, including multi-party computation (MPC)~\cite{goldreich1998secure}, secure enclaves, and fully homomorphic encryption (FHE)~\cite{Gentry09}.

The first realization of FHE was introduced by Gentry~\cite{Gentry09,GenPhD}, through the \emph{bootstrapping} mechanism that is vital for the wide applicability of FHE. Most FHE schemes rely on noise to ensure the security of the data, but the noise accumulates when homomorphic operations are applied. Bootstrapping reduces the noise of a ciphertext, therefore it allows more operations to be performed and frees the application developer to increase the size of encrypted programs.  

\subsection{Our Contribution} 

Our approach is based on FHE, which is designed to enable complex processing to be carried out on ciphertexts, without the need for any secret information. Therefore, untrusted servers can process private data. 

In this paper, we present a new method for Private Decision-Tree Evaluation (PDTE) with FHE. We restrict ourselves to secure \emph{inference}. While secure \emph{training}, using techniques such as Differential Privacy~\cite{dwork2006differential} or Federated Learning~\cite{bonawitz2019towards}, is also an interesting challenge in the PPML space, it is beyond the scope of our work. Our technique can easily be used with any tree-based model, and works for both regression and classification\footnote{We  implemented our method in \library{Concrete-ML} library~\cite{CML}}. We demonstrate our method on decision trees, random forests, and gradient boosted trees. 

Our contributions are the following:
\begin{enumerate}
    \item We use ciphertexts that store multi-bit integers and parameterize the crypto-system to allow for correct leveled accumulation of ciphertexts.
    \item We apply quantization on individual data features to control the message space required in the ciphertexts and, by leveraging tensorized computation on integers and programmable boostrapping (PBS), we implement PDTE in TFHE. 
    \item Crypto-system parameters are determined using an automated method relying on optimization. Thus it is possible to easily vary the size of the message space size, by adjusting the quantization of the input data, to find the best operating point that maximizes FHE inference speed while preserving accuracy.
\end{enumerate}

The plan of our paper is as follows: We first give an overview of FHE, \scheme{TFHE} and tree-based models. We then describe our method for PDTE with FHE. Next, we show our experimental results on a variety of datasets. We conclude with a discussion.

\subsection{Existing Works}

Boostrapping is a central feature of FHE, but not all implementations of machine learning in FHE  make use of it in practice. For example, schemes such as \scheme{CKKS}~\cite{CKKS17} uses approximate arithmetic, managing the noise budget in a way that avoids costly boostrapping by making the encrypted program contain only few multiplications. Similarly \scheme{BGV}~\cite{BGV12} or \scheme{BFV} use the same approach. Using only additions and multiplications for ML model inference has the drawback that models which use non-linear functions such as \operation{sign} or \operation{ReLU} have to rely on polynomial approximations of these functions. 

A number of methods have been proposed for PDTE with FHE. \cite{akavia2022privacy} relies on the polynomial approximation of the \operation{sign} function to compute the decisions in the trees. Since the degree of the polynomials must be kept low to stay within the noise budget, the approximation may have large error. To combat this issue, \cite{CryptoTree} expresses the decision tree as a neural network and fine-tunes it to adjust the decision thresholds. However, this adds complexity to the training process. \cite{tueno2020non} uses the BFV scheme which supports both an addition and a multiplication operation. They show an approach using bit-string representations and another with integer ciphertexts. Comparison is implemented using BFV multiplication between bits and decisions along a path are stored in a path vector. Finally, the decision for a leaf node is taken based on the multiplication of the decision bits along the path that leads to the leaf and the leaf's value. 

In the \scheme{TFHE} scheme ~\cite{CGGI16,CGGI17,CGGI20}, that is used in this work, any non-linear univariate function can be computed on ciphertexts by the \emph{programmable bootstrapping (PBS)}~\cite{CJP21} mechanism. A notable contribution to private decision tree evaluation, \cite{SortingHat} is based on TFHE, coupled with transciphering which helps bring down the ciphertext size. Their work represents inputs as strings of encrypted bits, corresponding to integers obtained by quantizing floating point data points using 11 bits of precision. This is in contrast to our approach that uses multi-bit ciphertexts. To perform classification \cite{SortingHat} uses homomorphic tree traversal computed using encrypted multiplication expressed through binary AND gates on individual bits. In our work we use parallel computation to evaluate all tree nodes using PBS, and to identify the decision path using a path code also using PBS. In their work, as in ours, the result of tree traversal is a one-hot vector containing a single value of one in the cell corresponding to the leaf that is chosen for a certain input. 

Other attempts, such as~\cite{meng2020privacy}, rely on additive homomorphic encryption~\cite{Pai99} and on order-preserving encryption ~\cite{OPE} which allows them to compare encrypted data with the same cost as comparing clear data. On the downside, such a combination of techniques may only satisfy a weaker security model than TFHE~\cite{boldyreva2011order} in the context where the server is not trusted. 

\section{Preliminaries}

\subsection{Fully Homomorphic Encryption}\label{sec:fhe}

An encryption scheme $f$ is said to have \emph{homomorphic} properties when there exist two operations $\cdot$ and $\sqcdot$ such that

    $$f(a \cdot b) = f(a) \sqcdot f(b),$$

\noindent for all valid messages $a$ and $b$. Often $\cdot$ and $\sqcdot$ are the same operations. Homomorphic schemes have been known since the introduction of public-key cryptography. Indeed $\scheme{RSA}$ has the multiplicative homomorphic property~\cite{RSA78}. Additive homomorphic schemes, such as \scheme{Paillier}, are also known~\cite{Pai99} and have been widely used, for example in voting schemes.

\emph{Fully homomorphic encryption} schemes are homomorphic for an universal set of operations. The field of cryptography had to wait until Gentry's breakthrough~\cite{Gentry09,GenPhD} to have a first construction of FHE. One key ingredient of his construction (and of those which followed) is the \emph{bootstrapping} mechanism, which allows to reduce the noise in a ciphertext. Several FHE schemes have since been introduced, which are both secure and practical, notably \scheme{BFV}~\cite{Bra12,FV12}, \scheme{GSW}~\cite{GSW13}, \scheme{BGV}~\cite{BGV12,BGV14}, \scheme{FHEW}~\cite{DM15}, \scheme{CKKS}~\cite{CKKS17}, and \scheme{TFHE}~\cite{CGGI16,CGGI17,CGGI20}.

%\paragraph{FHE and ML.}

%FHE has been identified as a great tool for privacy-enhancing technology, notably thanks to its security and the fact that it doesn't really change the protocol (as opposed to, for example, other techniques such as multi party computations). In the \scheme{TFHE} scheme, programmable bootstrapping (PBS) allows for the utilization of table lookup (TLU) during the bootstrapping process without incurring additional cost, as noted in~\cite{CJP21}. Thus, with PBS, it is possible to replace non-linear functions --- typically activations --- by TLUs.

\subsection{Integer Computations with \scheme{TFHE} and Corresponding Constraints}

\paragraph{Encoding multi-bit numbers}

\scheme{TFHE} supports both leveled operations, i.e. operations that increase noise, but also comes with fast boostrapping that is programmable. 

We use the approach from \cite{DiNN}, to represent integers with \scheme{TFHE}. To represent values that require $p$ bits, the torus is split into at least $2^p + 1$ slices, one for each possible message. To ensure leveled operations are always correct, $p$ must be the maximum bit-width obtained anywhere in the computation and the noise must be configured to take into account the clear constants that are multiplied with encrypted values. To correctly compute a dot-product such as $x \cdot w$ the standard deviation of the noise added during encryption of the plaintexts must be at most equal to the output noise standard deviation divided by $\left|w\right|_2$.

The PBS can perform a table look-up operation (TLU) on the input ciphertext, all the while reducing noise in the ciphertext. Furthermore, \scheme{TFHE} can be easily parameterized to ensure that noise accumulation due to leveled operations does not overwrite the message bits. 

Depending on crypto-system parameters, the PBS might have a probability of failure $p_{\texttt{error}}$ (usually chosen to be very small). The table look-up operation implemented with PBS can thus be defined as follows:

\begin{equation}
    TLU[x] = \begin{cases}
         T[x], \text{~with probability~}(1 - p_{\texttt{error}}) \\
         T[x + k], \text{~with probability~} < p_{\texttt{error}}, k = \{\pm 1, \pm 2, ...\}
    \end{cases}
\end{equation}

However, these properties come with some constraints ~\cite{Joy21}, and we summarize here those that impact our design for PDTE:

\begin{enumerate}
    \item Every input value and intermediate value within the model must be an integer type. To eliminate the possibility of noise corrupting the message during accumulation, our approach is to select crypto-system parameters that provision enough message and padding space to perform accumulation\footnote{Accumulating without noise corruption is detailed in ~\cite{CNAnnounce}}. Therefore, we cannot use sufficiently high precision to support floating point inputs and, thus, the training and test data must be quantized.
    
%    \item The maximum precision we can handle is 16 bits.\footnote{This limitation comes from the foundations of our dependency, namely the \library{Concrete-Library}~\cite{Concrete}} This is also true for all intermediate values (e.g., accumulators where the dot products are computed).

    \item Control-flow (i.e. branching) is not possible in any FHE scheme. We explain in Section~\ref{sec:cond} how to circumvent this limitation by leveraging the PBS, which is so far a tool only available in \scheme{TFHE}. In general, the PBS is well adapted to computing arbitrary univariate non-linear functions of encrypted inputs\footnote{The computational complexity of a PBS grows rapidly with the input bit-width and some implementations limit this bit-width to 16-bits.}. 
\end{enumerate}

\subsection{Decision Tree, Random Forest and XGBoost}

Tree-based models are a popular class of machine learning models that can perform both classification and regression. They are attractive as they are relatively interpretable~\cite{gilpin2018explaining}, easy to use (thanks to popular libraries such as \library{scikit-learn}), and remain to this day the state-of-the-art models when it comes to accuracy on tabular datasets~\cite{shwartz2022tabular}. Decision trees can be easily trained using a variety of algorithms, including the popular CART algorithm~\cite{breiman2017classification} which may train decision trees or ensembles of trees, such as Random Forests (RF) or Gradient Boosted Trees (XGBoost) ~\cite{CG16}. 

Once trained, tree-based models can be used for inference by traversing the trees from the root node to a leaf node which predicts a class or a regression value. Each internal node of the tree represents a test on an input feature and branches are followed according to the outcome of such tests. For ensembles one then applies a weighted sum of the individual tree outputs.

\subsection{Quantization}\label{sec:quant}

Quantization is the process of converting a continuous value into a discrete value. A common example of quantization is converting a signal from an analog domain to a digital domain. In the context of machine learning, quantization refers to the process of converting the learned parameters of a model but also to performing model inference using only quantized intermediary values. This can be achieved using a variety of methods, in particular uniform quantization~\cite{jacob2018quantization}. Quantization is often used to improve the efficiency of neural networks, both in terms of memory usage and computational speed. It can also be used to reduce the amount of data that needs to be stored and transmitted, which is important for applications such as mobile devices or embedded systems. 

\paragraph{Symmetric quantization.}
The straightforward approach is to use uniform quantization as follows:
\begin{equation}\label{eq:quant_symmetric}
q(x) = \operation{round}\left(\frac{x}{\Delta} \right)
\end{equation}
where $x$ is a real number, $q(x)$ is the \emph{quantized value} and $\Delta$ is the step size also called the \emph{scale}.

An appropriate step size takes the possible \operation{max} and \operation{min} of $x$ in consideration can be computed as follows:

\begin{equation}
\Delta = \frac{\operation{max}(x) - \operation{min}(x)}{2^{p}-1}
\label{eq:delta}
 \end{equation}
where $p$ is the number of bits that will be used to represent the quantized values. In this case, the quantized values will be integers between $-2^{p-1}$ and $2^{p-1}-1$. In this equation, we assume $\operation{max}(x)$ and $\operation{min}(x)$ are computed independently on a representative set of values of $x$. The value of $\Delta$ is used during the inference stage of the model through Equation \ref{eq:quant_symmetric}.

%These equations satisfy the following property:

%\begin{equation}\label{eq:0-0}
%    q(0) = 0
%\end{equation}

%\noindent or in other words, the quantization of the values zero in floating point is equal to zero once quantized. This is an important property when working with binary matrices or with sparse models (e.g. neural network subject to pruning).

\paragraph{Asymmetric quantization.}
Symmetric quantization works well as long as the distribution of the floating point values is symmetric around zero. In some cases, the distribution does not satisfy this property and it is more useful to use asymmetric quantization with a \emph{zero point} value $z$, defined as follows:

\begin{equation}\label{eq:quant_asymmetric}
q_a(x) = \text{round}\left(\frac{x}{\Delta}\right) + z
\end{equation}
The zero point is typically chosen such that the minimum of $x$ becomes the integer $0$ after quantization, i.e., $q_a(\operation{min}(x)) = 0$. Asymmetric quantization allows better use of the available representation bit-width.

\section{Private Decision Tree Evaluation with TFHE}
In this section, we present our technique for PDTE using FHE. As outlined in Section~\ref{sec:fhe}, \scheme{TFHE} imposes certain limitations and constraints, particularly in regard to condition and flow-operations.

Our method can be decomposed in three parts:

\begin{itemize}
\item training an integer tree-based model on quantized data, as detailed in Section~\ref{sec:meth_quant}. 
\item implementing integer tree-based models by replacing conditions with an evaluation of all branches and then retrieving the leaf node by looking up the decision path. Both steps are performed using both leveled operations and PBS
\item selecting crypto-system parameters based on analysis of the sequences of leveled operations and PBSs. The parameters are chosen so that leveled operations are always exact. PBS error level is configured by experimentation. 
\end{itemize}

\subsection{Quantizing Tree-Based Models}\label{sec:meth_quant}

In this section, we describe how we transform a floating-point model to an FHE-compatible integer one using quantization. 

Let $T = (V, E, L, \tau)$ be a decision tree with $V$ the set of nodes, $E$ the edges, $L$ the values in the tree leaves and $\tau$ the thresholds associated to the nodes. Let $N$ be the number of such trees in an ensemble. The following steps are applied to quantize a tree-based model:

\begin{enumerate}
    \item The training data $X$ is fully quantized and the model is trained in this new representation space with the classical training algorithms
    \item The decision thresholds are quantized. The training algorithms for tree-based models (e.g. \library{ginigain}, \library{entropy} or \library{xgboost}) choose decision thresholds between discrete feature values in the training data. 
    \item We apply either  \operation{ceil} or \operation{floor} to the thresholds to obtain quantized thresholds.
    \item The values stored in the leaves (either class weights or regression values) are quantized with asymmetric quantization. 
\end{enumerate}

\begin{algorithm}[H]
\caption{Quantization strategy}
\label{alg:quant_tree}
\textbf{Require}
\begin{itemize}
\item $X$: training dataset (in floating point)
\item $p$: the number of quantization bits
\end{itemize}

\textbf{Step 1:} $(X_q, Q_X) \gets \texttt{train\_quantizer}(X, p) $ \\
\textbf{Step 2:} $(V, E, \tau, L) \gets \operation{train}(X_q) $ \\
\textbf{Step 3:} $\tau' \gets \operation{floor}(\tau)$ \\
\textbf{Step 4:} $(L_q, Q_L) \gets \texttt{train\_quantizer}(L, p)$ \\
\textbf{Return:} $T = (V, E, L_q, \tau', Q_X, Q_L)$
\end{algorithm}

Asymmetric quantization is preferred for its greater precision, as it does not assume a symmetrical distribution of values around zero. In the first step it is used for quantizing input features. Let the $\texttt{train\_quantizer}(X, p)$ function be defined as the computation of $\Delta$ as in Eq.\ref{eq:delta} while setting $z = -\lfloor\frac{\operation{min}(x)}{\Delta}\rfloor$

Since a tree-based model does not perform linear combinations of the inputs, we can safely quantize every single feature independently of each other. This allows us to have a scale and zero point for each feature which is a great advantage when the input dimensions follow different distributions. In the case described here, the quantization zero-point can be ignored. Indeed the decision thresholds are learned over the quantized integer values representing the features. 

On the other hand, the quantization of the values stored in the leaves of the tree is applied globally for the entire vector of values. Thus, a single scale and zero-point are used for all of the leaf values. For ensemble models the leaf values will be represented by a matrix with of size $N \times m$ with $N$ being the number of trees in the ensemble and $m$ the number of leaf nodes.

%The quantization of operations in neural networks requires adherence to specific rules~\cite{jacob2018quantization}, making the process complex. However, in tree-based models, this complexity is avoided as the tree is learned over quantized input directly. As a result, the issue of propagating the scale and zero point through quantized operations to facilitate dequantization is not present in tree-based models.

\medskip

At this point, our FHE tree-based model is fully quantized. However, as we mentioned previously, control-flow operations are not directly possible in FHE, and require a different approach, which is explained in next section.

\subsection{Evaluating Decision Nodes in FHE}\label{sec:cond}

We evaluate all the decision nodes in the tree using PBS which can implement any univariate function, including comparison and equality test.

Consider a two-dimensional integer input space where each data point $x \in X$ belongs to the set $[0, 2^p)^2$. Here, $p$ represents the number of bits used to encode the features of $x$. Let the first feature of $x$ be denoted as $\feature{x}{1}$ and the second feature as $\feature{x}{2}$, with $p = 3$ in this example. Our goal is to classify each data point $x \in X$ into one of two classes, $C_0$ and $C_1$, by finding a function $f$ such that $x$ is assigned to $C_{f(x)}$ by our algorithm.

A simple boundary such as $\feature{x}{2} > 3$ can be represented by the decision stump (tree depth of 1) and expressed as follows:

\begin{equation*}
f(x) = \begin{cases} 0, & \text{~if~} x^{(2)} > 3 \\
                    1, & \text{otherwise}
      \end{cases}
\end{equation*}

This function can be implemented using a TLU as follows:

\begin{equation*}
f(x)=T[x], \text{ with } T = [1, 1, 1, 1, 0, 0, 0, 0]
\end{equation*}

\subsection{Tree Traversal Conversion to Tensor Operations}\label{sec:meth_tensor}

The tree traversal approach is a common way to perform tree model inference. However, in FHE, tree traversal is not possible since control-flow operations are not supported. To overcome this, we compute all branching decisions simultaneously by converting the tree traversal into tensor operations. Transforming a program that contains control-flow operations into one without branches is a common approach to accelerate computation over specific hardware (e.g. GPU). We follow the implementation of ~\cite{Hummingbird}.

Algorithm~\ref{alg:gemm} lists the tensors used in the conversion process as well as the process itself.

\begin{algorithm}[H]
\caption{GEMM Strategy for Tree Inference}
\label{alg:gemm}
\textbf{Require}
\begin{itemize}
\item $x_q$: quantized input example
\item $A$: feature-node selection matrix
\item $B$: threshold value
\item $C$: relationship between leaf nodes and internal nodes
\item $D$: count of internal nodes in the path from a leaf node to the tree root
\item $L_q$: mapping between leaf nodes and class labels
\end{itemize}

\textbf{Step 1:} $P \gets x_q \cdot A$ \\
\textbf{Step 2:} $Q \gets P < B$ \\
\textbf{Step 3:} $R \gets Q \cdot C$ \\
\textbf{Step 4:} $S \gets R == D$ \\
\textbf{Step 5:} $T \gets S \cdot L_q$ \\
\textbf{Return:} $T$
\end{algorithm}

$Q$ and $S$ are matrices of 1-bit integers, while $P$, $R$ and $T$ are matrices of multi-bit integers. $A, B, C, D, L_q$ are obtained by converting the model's tree structure. The matrices are created as follows.

\begin{enumerate}
    \item Tensor $A$ maps features to thresholds. For tree-ensembles $A$ is a 3-dimensional tensor and the mapping of different trees are stored along the 3rd dimension of $A$ :
\begin{equation*}
A_{kij} = \begin{cases} 1, & \text{~if threshold~} \tau_i \text{~is compared to feature ~} P_j \text{~in tree k~}\\
                    0, & \text{otherwise}
      \end{cases}
\end{equation*}

    \item Matrix $B$ contains the quantized thresholds:
\begin{equation*}
    B_{ki} = \tau'_i, \forall k
\end{equation*}

    \item Tensor $C$ is the tree-hierarchy encoder tensor.
\begin{equation*}
C_{kij} = \begin{cases} 

\begin{cases} 1, & \text{~if node ~} i \text{~is the left child~}  \\
                    -1, & \text{otherwise}
      \end{cases}
      
      & \text{~if node ~} i \text{~is a child of ~} j\\
                    0, & \text{otherwise}
      \end{cases}
\end{equation*}
    
    \item Tensor $D$ is used to identify the selected leaf node. Each leaf node in the graph can be reached by a decision path which can be encoded with a sequence of $[-1, 1]$ values, such as those in matrix $C$. Summing up those values along each decision path gives the path identification matrix $R$ in \textbf{Step 3}. Comparing $R$ with $D$ will uniquely identify the decision path, and thus the predicted leaf node, resulting in one-hot vectors for each tree. 
    
    \item Tensor $L_q$ encodes the values in the leaf nodes, as detailed in algorithm \ref{alg:quant_tree}. The matrix $S$ contains multiple one-hot vectors representing the predicted leaf nodes for each tree in the ensemble. The predicted output is obtained in \textbf{Step 5} as a dot product. 
\end{enumerate}

\subsection{Crypto-system Parameter Optimization}

The computation described in Algorithm \ref{alg:gemm} can be represented as a directed acyclic graph (DAG) containing accumulation operations and PBSs. We use the approach detailed in \cite{optimizer} to find crypto-system parameters, depending on the quantization bit-widths of the input values $X$ and of the constants in the $A$, $C$ and $L_q$ matrices. The 
$A$ constant matrix contains only 0/1 values so it is represented with single bits. Multiplying it with the $X$ vector produces a result $Q$ with the same bit-width as $X$. Next, matrix $R$ contains values that are at most equal to the tree depth $d$, which is usually quite low (5-10), requiring at most $\text{log}_2(d) + 1$ bits. The final step is a dot product between two-dimensional matrices $S$ and $L_q$.

\begin{figure}
    \centering
    \includegraphics[width=0.8\textwidth]{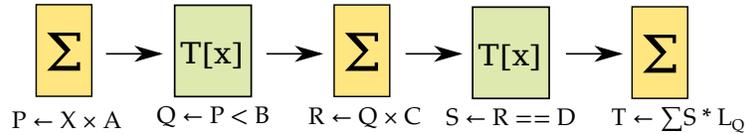}
    \caption{TFHE operations representing algorithm \ref{alg:gemm}. The $\Sigma$ nodes represent leveled accumulation and $T[x]$ nodes represent table look-up operations implemented with PBS }
    \label{fig:enter-label}
\end{figure}

The final step, $S \leftarrow R \cdot L_q$ can be split in two. Indeed the leaf values for each tree are quantized to the same bit-width as the input. However, when multiple trees are present in the ensemble, their predictions need to be summed as well. This second sum can be performed in the clear, to avoid accumulating a large bit-width value that would exceed the bit-width of the input. 

\begin{itemize}
    \item \textbf{Step 5.1 (encrypted): } $T_k \leftarrow \sum_i S \times L_q $ 
    \item \textbf{Step 5.2 (clear): } $T \leftarrow \sum_k T_k $
    
\end{itemize}

Applying this modification to Algorithm \ref{alg:gemm} upper bounds the bit-widths of the integers in the entire computation graph to the input bit-width $p$ plus one. The additional bit is required to represent signed values in matrix $R$, since the input features are quantized to unsigned integers on $p$ bits.

When a tree-ensemble model is used, the vector of results of the individual trees must be aggregated over the trees. Next, for classification, the \operation{argmax} will need to be computed over the decisions. We perform the aggregation as shown in \textbf{Step 5.1 and 5.2}, and the \operation{argmax} is computed on the client side, after decryption. Moreover, the client will apply the inverse operation to quantization, dequantization, to get back the values that would have been predicted by the floating point algorithms without encryption.

An optimization algorithm divides the graph into subgraphs containing leveled operations (addition of encrypted values, multiplication of encrypted values with clear values) and ending with a PBS. Since the clear values are known (the matrices $A, B, C, D, L_q$ are constant), the maximal bit-width of accumulated values can be computed. In our case the maximum bit-width will be $p+1$. It is therefore possible to generate crypto-parameters that provision sufficient message-space for any input $x_q$, such that leveled operations are always correct. The values of $x_q$ are assumed to always be bounded by the input bit-width $p$ chosen during model training.

\section{Experimental Results}\label{sec:expe}

In this section we describe the experiments we performed to evaluate our method and we discuss the results. We first test the accuracy of quantized models, as defined in Section~\ref{sec:meth_quant}, and we compare it with the one of floating point (FP32) models. 

PDTE has the drawback of having a significantly longer inference time than the clear model inference. This execution time depends mainly on two factors: the crypto-system parameters and the complexity of the FHE circuit. As the security level is constant,\footnote{Currently in \library{Concrete-ML}, security level is set to  128-bit.} cryptographic parameters depend mainly on the precision we use (i.e., the upper value of the bit width of intermediate values) while the complexity of the FHE circuit is directly correlated the model hyper-parameters. For tree models these are: the total number of nodes, the maximum depth $d$ and the number of trees in an ensemble, $N$.

We performed experiments with three different types of tree-based model for classification, as they are the most common~\cite{banerjee2019tree}:
\begin{itemize}
    \item DecisionTreeClassifier from \library{scikit-learn} library.
    \item RandomForestClassifier from \library{scikit-learn} library.
    \item XGBoostClassifier from the \library{xgboost} library~\cite{CG16}.
\end{itemize}

%\subsection{Implementation in \library{Concrete-ML}}

%In \library{Concrete-ML}, the flow has been made easy for any data-scientist to have their usual tasks remain unchanged. User can use the API to train models and predict similarly as in the \library{scikit-learn}. In summary, the process of converting the model to its FHE equivalent involves the following steps.

%A tree-based model is trained over quantized data using one of the parent library (namely, \library{scikit-learn} or \library{xgboost}). The model is then exported to ONNX using \library{hummingbird}, decision thresholds are converted to integer and prediction values (in terminal leaves) are quantized. Both input and output quantizer remain at the user's disposal as they are needed to pre/post process the data before and after FHE execution. The ONNX model is then converted to \library{Numpy} functions and given to \library{Concrete-Numpy} for compilation and bounds measurements. An optimizer is run to compute the optimal cryptographic security parameters (a necessary information for public/private key generation). The FHE binary is produced by \library{Concrete-Compiler} that implements every FHE operations.

%Once the model is trained and compiled, the user can use \library{Concrete-ML} Python API to encrypt, quantize/dequantize and run the FHE execution easily. One can refer to our~\cite{SpamTuto} available in our repository, where we handle Spam detection in FHE on a real dataset.

\subsection{Quantization Precision And FHE Inference Time.}

\begin{figure}[h]
\centering
    \includegraphics[width=0.45\textwidth]{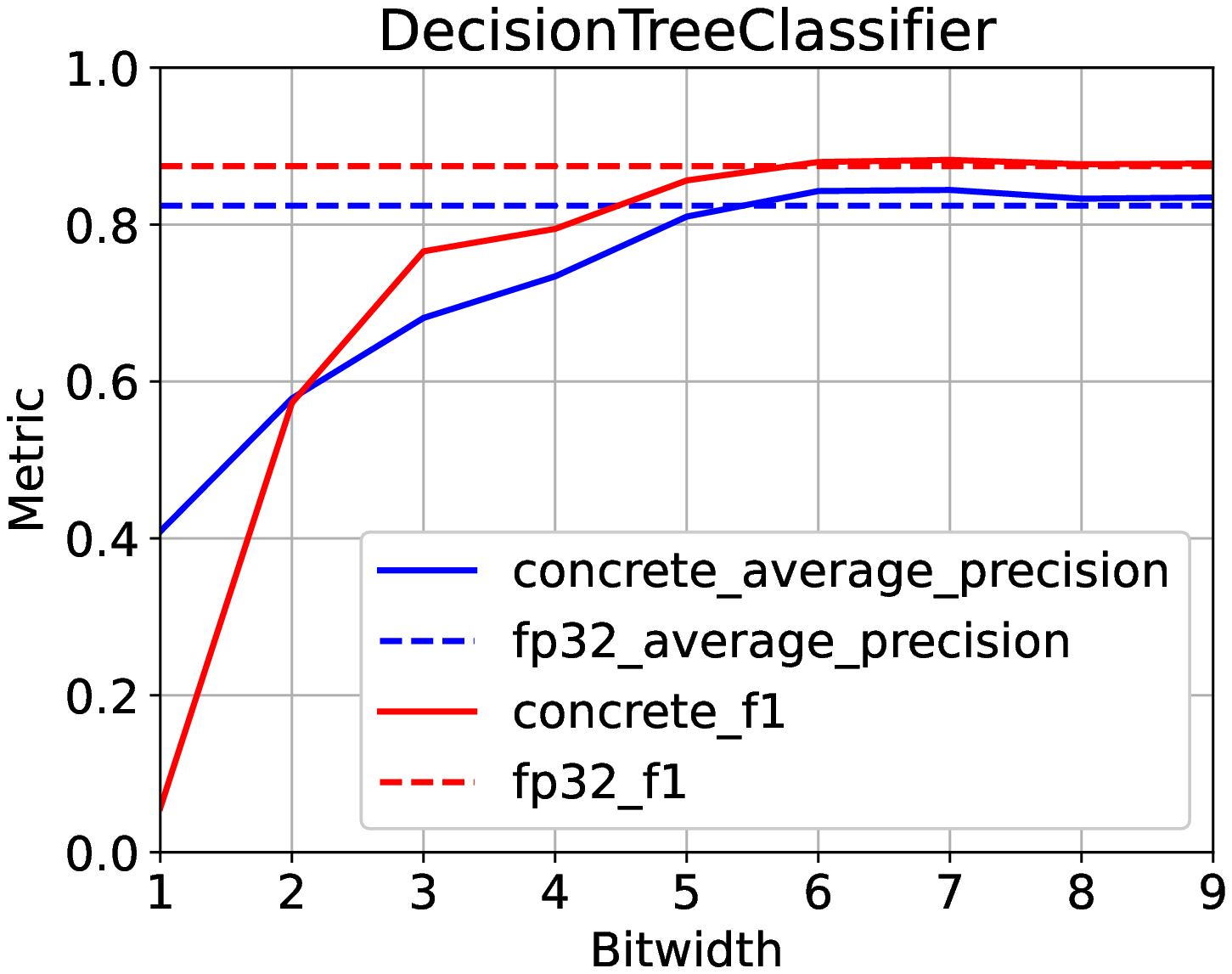}
\includegraphics[width=0.45\textwidth]{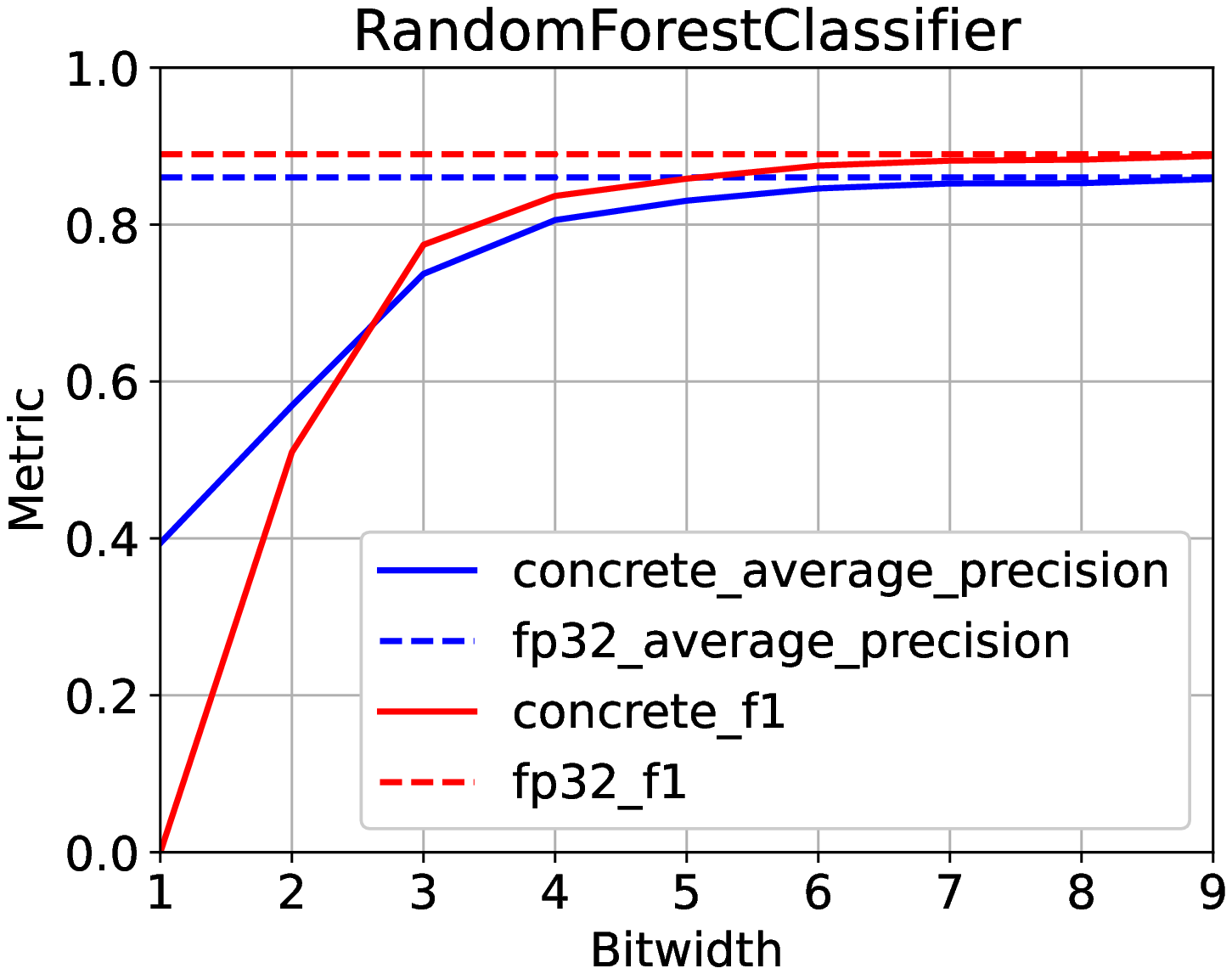}
\includegraphics[width=0.45\textwidth]{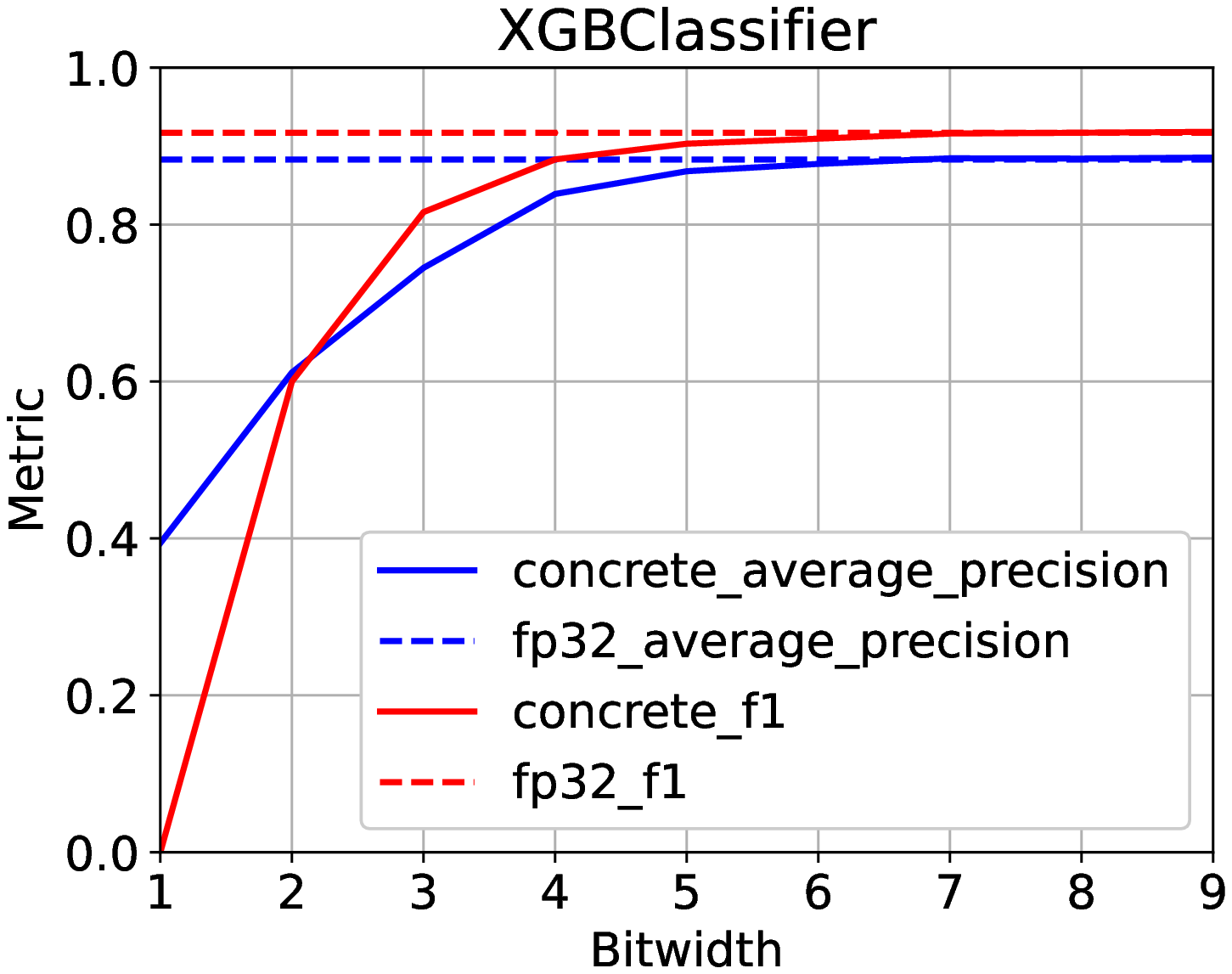}
\caption{Experiment reporting the f1-score and average precision with varying precision on the spambase dataset.}
\label{fig:spambase_concrete_bit_width_accuracy}
\end{figure}

First we study the impact of quantization on model accuracy. In Figure~\ref{fig:spambase_concrete_bit_width_accuracy}, we compute the \operation{f1} score and average precision (\operation{AP}) for different quantization bit-widths and we plot them along with the metrics from the floating point value model. For the three different tree-based models, we can observe a convergence of the quantized model toward the full precision model. In the case of decision trees the accuracy with quantized data can sometimes exceed the one obtained by training on floating point data since quantization regularizes the training procedure, avoiding over-fit. Ensemble models are more robust to over-fit and do not show this behavior. 

% \begin{figure}[htp]
% \centering
% \begin{minipage}{.45\textwidth}
%   \centering
%   \includegraphics[width=\linewidth]{figures/DecisionTreeClassifier}
% \end{minipage}%
% \begin{minipage}{.45\textwidth}
%   \centering
%   \includegraphics[width=\linewidth]{figures/RandomForestClassifier}
% \end{minipage}

% \medskip
% \begin{minipage}{.45\textwidth}
%     \centering
%     \includegraphics[width=\linewidth]{figures/XGBClassifier}
% \end{minipage}
% \caption{Experiment reporting the f1-score and average precision with varying precision on the spambase dataset.}
% \label{fig:spambase_concrete_bit_width_accuracy}
% \end{figure}

\begin{figure}[h]
\centering
\includegraphics[width=0.5\textwidth]{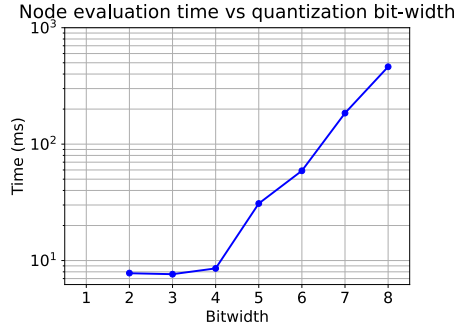}
\caption{FHE node evaluation latency for decision nodes with $p_{\texttt{error}} = 10^{-40}$. Decisions are batched and computed in parallel on 8 CPU cores. }
\label{fig:spambase_bitidth_inference_time}
\end{figure}

Such a behavior is expected  and the natural choice would be obviously to select the highest bit width. However, FHE execution time is impacted by the bit-width. To have a better understanding of the impact of quantization, we run an experiment in Figure~\ref{fig:spambase_bitidth_inference_time} where we compute the FHE inference time for the three models at different quantization bit-widths. This figure shows that computation time for a single decision node (one PBS) increases exponentially for $p > 4$.

Figures~\ref{fig:spambase_concrete_bit_width_accuracy} and~\ref{fig:spambase_bitidth_inference_time} provide a good overview of the trade-off between model accuracy vs. FHE inference time. We can see that a large increase in FHE inference time is found to be between starting at 7 bits. On the other hand, 5 and 6 bits give results close to those of the FP32 model: a less than 2 percentage point drop in the metrics reported for 6 bits precision. In the following section, we use $p=6$ bits as our quantization precision for both the input features and output tree values.

\subsection{Experiments on Various Datasets}

Table~\ref{tab:experiments_datasets} presents the mean results of 5-fold cross validation repeated 3 times (or 15 runs in total per model per dataset).

Hyper-parameters used for both the FHE and sklearn models are:
\begin{itemize}
    \item $N$, the number of estimators, is set to 50 and defines the number of trees in the ensemble for both Random Forest (RF) and Gradient Boosting (XGB).
    \item maximum depth $d$ is set to 5 for both the decision tree (DT) and XGB. Random forest is left unbounded such that the tree can fully expand. This parameter defines the maximum depth a tree can have once fully trained.
    \item the maximum number of nodes \#nodes is left unbounded and is measured once the tree is trained
\end{itemize}

\begin{table}[]
    \centering
\begin{tabular}{llllllrl}
\toprule
              dataset        &         & accuracy &      f1 &      AP & \#nodes &  Latency (s) & FHE/Clear ratio \\
\midrule
spambase (\#features: 57) & FHE-DT &    91.0\% &   88.0\% &   84.3\% &    23 &     1.807 &            904x \\
                      & FP32-DT &    90.3\% &   87.4\% &   82.4\% &   - &     0.002 &                 \\
                      \cline{2-8}
                      & FHE-XGB &    93.1\% &   90.9\% &   87.7\% &   350 &    17.770 &           5923x \\
                      & FP32-XGB &    93.6\% &   91.7\% &   88.3\% &   - &     0.003 &                 \\
                      \cline{2-8}
                      & FHE-RF &    90.9\% &   87.5\% &   84.6\% &   750 &    35.377 &          11792x \\
                      & FP32-RF &    91.8\% &   89.0\% &   86.0\% &   - &     0.003 &                 \\
\midrule
wine (\#features: 13) & FHE-DT &    90.8\% &       - &       - &     7 &     1.071 &           1071x \\
                      & FP32-DT &    90.5\% &       - &       - &   - &     0.001 &                 \\
                      \cline{2-8}
                      & FHE-XGB &    96.4\% &       - &       - &   900 &    37.338 &          37338x \\
                      & FP32-XGB &    95.9\% &       - &       - &   - &     0.001 &                 \\
                      \cline{2-8}                      
                      & FHE-RF &    98.5\% &       - &       - &   500 &    21.330 &          21330x \\
                      & FP32-RF &    98.1\% &       - &       - &   - &     0.001 &                 \\
\midrule
heart-h (\#features: 13) & FHE-DT &    61.0\% &       - &       - &    21 &     1.552 &           1552x \\
                      & FP32-DT &    60.0\% &       - &       - &   - &     0.001 &                 \\
                      \cline{2-8}
                      
                      & FHE-XGB &    66.8\% &       - &       - &  1750 &    81.938 &          40969x \\
                      
                      & FP32-XGB &    65.5\% &       - &       - &   - &     0.002 &                 \\
                      \cline{2-8}
                      
                      & FHE-RF &    66.8\% &       - &       - &   750 &    34.688 &          34688x \\
                      & FP32-RF &    66.4\% &       - &       - &   - &     0.001 &                 \\
\midrule
wdbc (\#features: 30) & FHE-DT &    94.2\% &   92.0\% &   88.4\% &    15 &     1.180 &            590x \\
                      & FP32-DT &    93.9\% &   91.7\% &   87.3\% &   - &     0.002 &                 \\
                      \cline{2-8}
                      
                      & FHE-XGB &    96.5\% &   95.1\% &   92.8\% &   350 &    17.124 &           8562x \\
                      & FP32-XGB &    96.4\% &   94.9\% &   92.6\% &   - &     0.002 &                 \\
                      \cline{2-8}
                      
                      & FHE-RF &    95.6\% &   93.9\% &   91.2\% &   700 &    30.744 &          15372x \\
                      & FP32-RF &    95.3\% &   93.4\% &   90.4\% &   - &     0.002 &                 \\
\midrule
adult (\#features: 14) & FHE-DT &    83.6\% &   60.4\% &   50.3\% &    30 &     1.892 &           1892x \\
                      & FP32-DT &    83.6\% &   60.4\% &   50.3\% &   - &     0.001 &                 \\
                      \cline{2-8}
                      
                      & FHE-XGB &    84.8\% &   64.9\% &   53.8\% &   350 &    18.551 &          18551x \\
                      & FP32-XGB &    84.8\% &   65.2\% &   53.9\% &   - &     0.001 &                 \\
                      \cline{2-8}
                      
                      & FHE-RF &    83.4\% &   57.6\% &   49.2\% &   750 &    37.494 &          18747x \\
                      & FP32-RF &    83.4\% &   57.6\% &   49.2\% &   - &     0.002 &                 \\
\midrule
steel (\#features: 33) & FHE-DT &    97.2\% &   96.1\% &   92.5\% &     5 &     0.954 &            477x \\
                      & FP32-DT &    97.2\% &   96.1\% &   92.5\% &   - &     0.002 &                 \\
                      \cline{2-8}
                      
                      & FHE-XGB &   100.0\% &  100.0\% &  100.0\% &   150 &     8.778 &           4389x \\
                      & FP32-XGB &   100.0\% &  100.0\% &  100.0\% &   - &     0.002 &                 \\
                      \cline{2-8}                      
                      & FHE-RF &    96.9\% &   95.4\% &   93.6\% &   700 &    27.047 &          13524x \\
                      & FP32-RF &    95.9\% &   93.9\% &   91.4\% &   - &     0.002 &                 \\
\bottomrule
\end{tabular}

    \caption{FHE vs FP32 tree-based experiments. The accuracy, \operation{f1}-score and average precision (\operation{AP}) are averaged over 15 runs. The inference time per model is reported in the Latency column and the execution time ratio between FHE and FP32 model is computed. The PBS probability of error is set to $p_{\texttt{error}} = 10^{-40}$. We do not report \operation{f1}-score and \operation{AP} for multi-class datasets.}
    \label{tab:experiments_datasets}
\end{table}

%\textbf{Accuracy} between FP32 and FHE is closely matched for every dataset. The wine dataset shows the largest variance in all metrics. This can be caused by its limited sample size of only 178 examples which makes it challenging properly represent with quantization as only a fraction of these examples are presented to the model for each run. This is reflected by the higher standard deviation seen in all models and metrics for this particular dataset. For optimal quantization outcomes, input features should exhibit a balanced distribution and have a proper representation of the data in the training set.

The experiment shows interesting properties of the tree-based models. Decision trees, trained with a limited depth $d=5$, have the fastest average FHE execution time. However, due to their lower capacity, their accuracy is lower than the one of ensemble methods. The ratio of latency in FHE versus FP32 is roughly 10-20000x on average for ensemble methods which show the closest accuracy with respect to the FP32 models.

\begin{figure}[h]
\centering
    \includegraphics[width=0.45\textwidth]{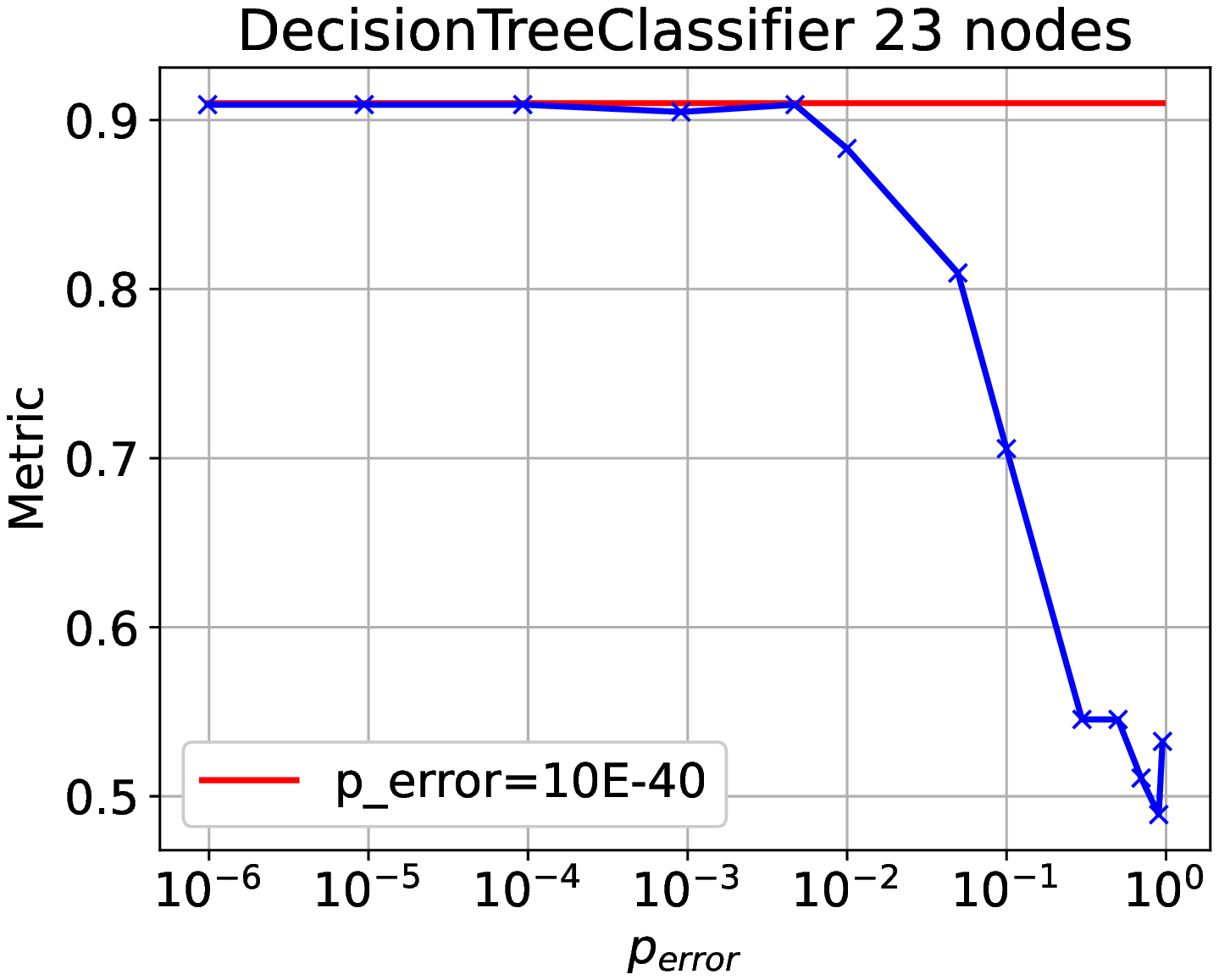}
\includegraphics[width=0.45\textwidth]{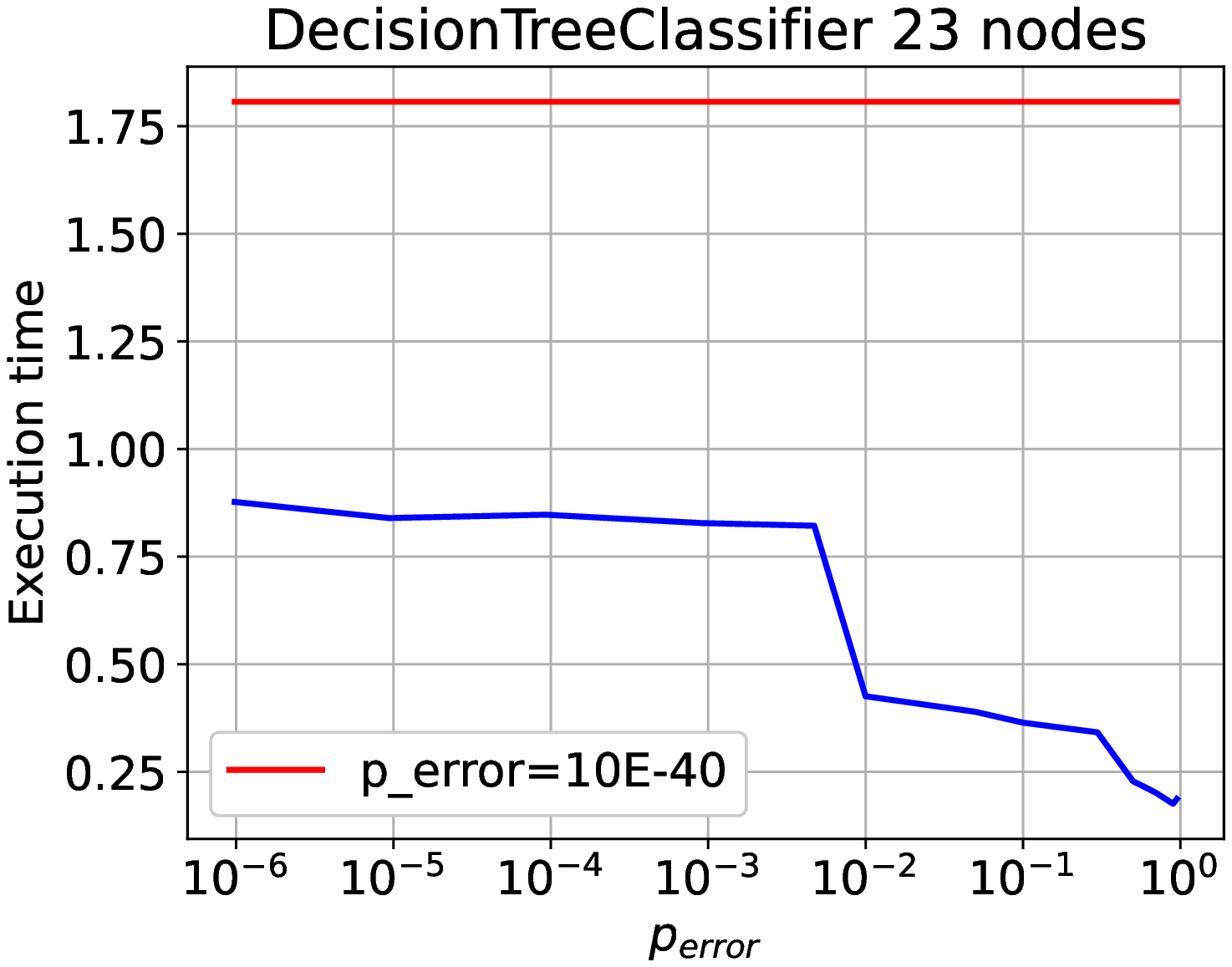}
\caption{Varying the probability of PBS error}
\label{fig:spam_pbs_error}
\end{figure}

Next, we performed experiments on the impact of the $p_{error}$ probability of error of the PBS. Figure \ref{fig:spam_pbs_error} shows a grid-search for the best $p_{error}$ value that maintains accuracy while minimizing latency. A value as high as $p_{error} = 0.05$ maintains accuracy while providing an 8x speed-up. 

Based on the experiments above, we configured decision tree models with 6 bits of precision and a $p_{error} = 0.05$. We performed experiments on trees with the same structure as those in \cite{BFVTree}  and compared results.

\begin{table}[]
\centering
\begin{tabular}{cccccccc}
\hline
Dataset & 
sklearn acc & 
\multicolumn{1}{p{2cm}}{\centering  Concrete-ML \\ acc} &
$d$ & 
\#nodes & 
\multicolumn{1}{p{2cm}}{\centering Latency (s) \\ $p_{error} = 10^{-40}$} & 
\multicolumn{1}{p{2cm}}{\centering Latency (s) \\  $p_{error} = 0.05$}& 
\multicolumn{1}{p{2cm}}{\centering \cite{BFVTree} \\ latency (s)} \\ \hline
spam    & 93,0\%           & 95,0\%               & 13    & 57              & 3.36            &  0.62
 & 3.66          \\
heart   & 61,0\%           & 56,0\%               & 3     & 4               & 0.84            &  0.22
 & 0.94          \\ \hline
\end{tabular}
\caption{Comparison with a \scheme{BFV}-scheme based implementation of PDTE.}
\end{table}

\section{Conclusion}

The present study offers a method for the conversion of tree-based models into their fully homomorphic encryption (FHE) equivalent, thus providing a secure mechanism for the deployment of machine learning models by service providers.

Our solution offers numerous advantages. To the best of our knowledge, it is the first tree-based solution to offer (i) easy customization that can obtain an optimal quantization / inference speed trade-off, (ii) compatibility with ensemble methods, and (iii) accuracy equivalent to execution in the clear.

Security is directly handled under the hood: crypto-system parameters are determined automatically and are optimal for a specific tree model, ensuring correctness of the leveled computations. 

We believe this method represents a significant advancement in the field and the method, as implemented in the open-source library \library{Concrete-ML}, is a step forward to democratizing privacy-preserving machine learning. 

\bibliographystyle{alpha}
\bibliography{biblio}

\end{document}